\documentclass[A4paper,unsortedaddress,superscriptaddress,prmaterials,twocolumn]{revtex4-2}
\usepackage[breaklinks=true]{hyperref}
\usepackage{epsfig}
\usepackage{graphicx}
\usepackage{subfigure}
\usepackage{latexsym}
\usepackage{color}
\usepackage{fullpage}
\usepackage{dcolumn}
\usepackage{bm}
\usepackage{ulem}
\usepackage{units}
\usepackage{amsmath}

\begin{document}

\title{Structure of graphene and a surface carbide grown on the (0001) surface of rhenium}

\author{Estelle Mazaleyrat}
\affiliation{Univ. Grenoble Alpes, CEA, IRIG/DEPHY/PHELIQS, 38000 Grenoble, France}
\affiliation{Univ. Grenoble Alpes, CNRS, Grenoble INP, Institut NEEL, 38000 Grenoble, France}
\author{Monica Pozzo}
\affiliation{Department of Earth Sciences, London Centre for Nanotechnology and Thomas Young Centre at University College London, University College London, Gower Street, London WC1E 6BT, United Kingdom}
\author{Dario Alf\`{e}}
\affiliation{Department of Earth Sciences, London Centre for Nanotechnology and Thomas Young Centre at University College London, University College London, Gower Street, London WC1E 6BT, United Kingdom}
\affiliation{Dipartimento di Fisica Ettore Pancini, Universit\`a di Napoli Federico II, Monte Sant'Angelo, I-80126 Napoli, Italy}
\author{Alexandre Artaud}
\affiliation{Univ. Grenoble Alpes, CEA, IRIG/DEPHY/PHELIQS, 38000 Grenoble, France}
\affiliation{Univ. Grenoble Alpes, CNRS, Grenoble INP, Institut NEEL, 38000 Grenoble, France}
\author{Ana Cristina G\'{o}mez Herrero}
\affiliation{Univ. Grenoble Alpes, CEA, IRIG/DEPHY/PHELIQS, 38000 Grenoble, France}
\author{Simone Lisi}
\affiliation{Univ. Grenoble Alpes, CEA, IRIG/DEPHY/PHELIQS, 38000 Grenoble, France}
\author{Val\'erie Guisset}
\affiliation{Univ. Grenoble Alpes, CEA, IRIG/DEPHY/PHELIQS, 38000 Grenoble, France}
\author{Philippe David}
\affiliation{Univ. Grenoble Alpes, CEA, IRIG/DEPHY/PHELIQS, 38000 Grenoble, France}
\author{Claude Chapelier}
\affiliation{Univ. Grenoble Alpes, CNRS, Grenoble INP, Institut NEEL, 38000 Grenoble, France}
\author{Johann Coraux}
\email{johann.coraux@neel.cnrs.fr}
\affiliation{Univ. Grenoble Alpes, CEA, IRIG/DEPHY/PHELIQS, 38000 Grenoble, France}

\begin{abstract}
Transition metal surfaces catalyse a broad range of thermally-activated reactions involving carbon-containing-species -- from atomic carbon to small hydrocarbons or organic molecules, and polymers. These reactions yield well-separated phases, for instance graphene and the metal surface, or, on the contrary, alloyed phases, such as metal carbides. Here, we investigate carbon phases on a rhenium (0001) surface, where the former kind of phase can transform into the latter. We find that this transformation occurs with increasing annealing time, which is hence not suitable to increase the quality of graphene. Our scanning tunneling spectroscopy and reflection high-energy electron diffraction analysis reveal that repeated short annealing cycles are best suited to increase the lateral extension of the structurally coherent graphene domains. Using the same techniques and with the support of density functional theory calculations, we next unveil, in real space, the symmetry of the many variants (two six-fold families) of a rhenium surface carbide observed with diffraction since the 1970s, and finally propose models of the atomic details. One of these models, which nicely matches the microscopy observations, consists of parallel rows of eight aligned carbon trimers with a so-called $(7\times\sqrt{\mathrm{19}})$ unit cell with respect to Re(0001).

\end{abstract}


\maketitle

\section{\label{intro}Introduction}

Graphene growth on strongly interacting metallic substrates such as Re(0001) \cite{miniussi2011, miniussi2014, tonnoir2013,papagno2013,gao2017}, Ru(0001) \cite{wang2008,marchini2007,sutter2008,sutter2009-Ru}, Rh(111) \cite{wang2010} and Ni(111) \cite{gamo1997} has the advantage of selecting one crystallographic orientation of graphene with respect to its substrate. This is related with a tendency to form covalent bonds between graphene and the substrate \cite{wang2010,gao2017}, as opposed to graphene grown on weakly interacting metallic substrates such as Ir(111) \cite{pletikosic2009,busse2011}, Pt(111) \cite{zi-pu1987,land1992,sutter2009-Pt}, Cu(111) \cite{gao2010}, Ag(111) \cite{kiraly2013} and Au(111) \cite{nie2012}, where van der Waals bonding is dominant \cite{busse2011,wang2010,gao2017}. Graphene growth on the latter kind of substrates results in domains with a large number of possible crystallographic orientations with respect to the substrate \cite{zi-pu1987,land1992,gao2010,merino2011,loginova2009,hattab2011,sutter2009-Pt,artaud2016}, which is problematic in the prospect of the production of highly ordered graphene. 

After its growth with a well-defined crystallographic orientation on strongly interacting metallic substrates, single-layer graphene has been successfully transferred to other, application-ready substrates~\cite{ago2010,sutter2010,iwasaki2010,koren2013}. Despite these successes, a number of difficulties associated with the use of these substrates must be borne in mind, since they are not easily overcome. In particular, on Rh(111), Ni(111) and Re(0001) surfaces, there is a subtle competition between the formation of different carbon phases, graphene and metal carbides \cite{gall1987,miniussi2014,dong2012,qi2017,lahiri2011}. The relative stability of these phases depends on the temperature. In addition, even when in some temperature range, graphene is the most stable species on the surface, its nucleation may require a high carbon chemical potential (a high carbon adatom concentration) \cite{dong2012}, without which the other carbon phases exclusively form on the surface. In the case of Re(0001), which is at the same time a growth substrate and a platform to induce superconductivity up to 2.1~K in graphene by a proximity effect \cite{tonnoir2013} (the strong graphene-Re interaction destroying the unique electronic spectrum of graphene can be suppressed while maintaining the proximity effect using Au intercalation \cite{mazaleyrat2020}), selective growth of graphene is limited to a narrow window of growth parameters (substrate temperature, hydrocarbon gas pressure and exposure time) \cite{miniussi2014}.

Although obviously considered as detrimental in the context of graphene production, the surface metal carbides that sometimes grow at the expense of graphene provide well-ordered surface structures with crystallographic unit cells at the scale of nanometers. This makes them efficient nanopatterns driving the self-organization of functional clusters, especially metallic clusters of interest for magnetic data storage or heterogeneous catalysis applications \cite{bansmann1999,varykhalov2005,varykhalov2008}.


Only few works devoted to surface carbides formed on the (0001) surface of rhenium have been reported so far \cite{zimmer1972,miniussi2014,qi2017}. An ordered carbide overlayer was previously identified with low-energy electron diffraction (LEED) \cite{zimmer1972,miniussi2014}, from which unit cell vectors of $7\vec{a}_{Re,1}$ and $2\vec{a}_{Re,1}+5\vec{a}_{Re,2}$ (with $\vec{a}_{Re,1}$, $\vec{a}_{Re,2}$ the Re(0001) unit vectors) were inferred. In the extended Wood's notation \cite{artaud2016}, this superstructure corresponds to $(7R0\times\sqrt{19}R(-23.4))$, and is referred to as (7$\times$$\sqrt{19}$) \cite{zimmer1972,miniussi2014}. It was never observed directly with a local probe method.

Here, we investigate the structure of graphene and the (7$\times$$\sqrt{19}$) surface carbide grown on Re(0001). To characterise lattice parameters and learn about the symmetry of these carbon phases, we used an ensemble-averaging \textit{in operando} probe, reflection high-energy electron diffraction (RHEED). A local probe analysis using scanning tunneling microscopy (STM) is less suited for this purpose, as it reveals, for instance, slight spatial structural variations about the average structure of graphene/Re(0001) \cite{tonnoir2013,artaud2016}. However, STM was used to get insights into the graphene and carbide morphologies and to resolve the atomic details of the carbide with the help of density functional theory (DFT) calculations. We find that increasing the number of annealing cycles during graphene growth increases the quality and extension of graphene domains. As it is known, excessive annealing temperatures promote the formation of the (7$\times$$\sqrt{19}$) surface carbide, which we observed for the first time by STM. The carbide has various crystallographic orientations on the surface, and exists in the form of six kinds of domains, each with a specific atomic arrangement. Four atomic models, complying with former X-ray photoelectron spectroscopy (XPS) characterisation, are proposed for the carbide structure of one of these domains, whose relative stability is probed with DFT calculations. The models are low-density C phases (considering the topmost Re layer, a Re$_{11}$C$_{8}$ compound) compared to graphene/Re(0001); they consist of a layer of parallel rows of C trimers attached to the hollow binding sites of the substrate, which experiences significant lattice distortions in its topmost layers.

\section{\label{methods}Materials and methods}

The surface preparation of Re(0001) single crystals substrate [see details in the Supplementary Material (SM)] and the subsequent growth of graphene and of the carbide were performed under ultrahigh vacuum (UHV), in a chamber having a base pressure of 10$^{-10}$~mbar.

Graphene was grown using the method described in Ref.~\onlinecite{miniussi2014}, which consists in a succession of rapid annealing cycles under C$_2$H$_4$ atmosphere (C$_2$H$_4$ introduced in the UHV chamber \textit{via} a leak-valve). The Re(0001) surface was first exposed to an ethylene pressure $P_\mathrm{C_2H_4}$ of 1 to 5$\cdot$10$^{-7}$~mbar, at room temperature. This is expectedly a way to reach a high-enough carbon surface concentration for graphene nucleation after the temperature is increased in the next step (exposing the surface to C$_2$H$_4$ only after the temperature has been increased does not allow to form graphene), similar to the case of graphene/Rh(111) \cite{dong2012}. The temperature was next rapidly increased to 870-970~K (measured with a pyrometer assuming an emissivity of 0.3), and then slowly decreased to 520~K. This cycle was repeated several times. During this process, it is expected that the graphene coverage approaches 100\% already after the few first cycles, since the ethylene precursor is provided at the metal surface held at elevated temperatures. The cycling process is thought to increase the structural quality at graphene thanks to the high temperature (870-970~K) treatment, while limiting the decomposition of graphene \textit{via} C-C bond breaking that would be promoted by increasing time spent at elevated temperature \cite{miniussi2014}. After the temperature cycles under ethylene partial pressure, a final annealing cycle was performed in absence of C$_2$H$_4$. Depending on the duration of this last step, part of the graphene was transformed into rhenium carbide.

During the exposure to C$_2$H$_4$ and thermal treatment, snapshots of the RHEED patterns (10~keV electron source) produced by the surface were acquired at a rate of typically 1~Hz using a CCD camera (1280$\times$960 pixels). The structural analysis consisted in extracting the position and full-width at half maximum (FWHM) of the diffraction streaks in the RHEED patterns, informing on the average lattice parameter and sizes of the corresponding structurally-coherent domains, respectively. The extraction relied on fits of the experimental streak profiles extracted within the rectangular regions of interest displayed in Fig.~\ref{fig1}(a), using a set of Lorentzians. More details are given in the SM, Fig.~S2 in particular \cite{SM}.

STM measurements were performed at room temperature before and after one or several temperature cycles with the samples transferred under UHV to a commercial Omicron UHV-STM 1, with W chemically-etched tips.

Calculations were performed using DFT~\cite{hohenberg1964,kohn1965}, with exchange-correlation effects included at the level of the PBE-GGA~\cite{perdew1996} functional. We used the projector augmented wave~\cite{blochl1994} (PAW) method as implemented in VASP~\cite{kresse1996,kresse1999} to account for the core electrons of both Re and C atoms, with the 6$s$ and 5$d$ electrons of Re and the 2$s$ and 2$p$ electrons of C explicitly included in the valence band. To represent the Kohn-Sham orbitals we used a plane-wave expansion with a kinetic energy cutoff of 400~eV.  As the calculations are fully periodic in the three spatial directions, surfaces have been represented in the usual slab geometry, with 4 layers of Re and the C atoms adsorbed on the top layer. The bottom two layers of Re were kept fixed to their bulk geometry, and the other atoms in the simulation cell were fully relaxed. The surface geometry was defined using a supercell having the shape of a parallelogram, with unit vectors $7 \vec{a}_{Re,1}$, $2 \vec{a}_{Re,1} + 5 \vec{a}_{Re,2}$ and $10 \vec{a}_{Re,3}$ (with $\vec{a}_{Re,1,2,3}$ the Re(0001) unit vectors). The length of $10\vec{a}_{Re,3}$ insured a vertical distance of about 20~\AA\; between the surface and the bottom layer of the periodic image of the slab, which is large enough to decouple the two, as demostrated by tests using a longer $\vec{a}_3=a\times(0,0,15)$, which showed essentially no energy difference.  Brillouin-zone sampling was performed using a 2$\times$3$\times$1 Mokhorst-Pack~\cite{monkhorst1976} grid. Tests with a 3$\times$5$\times$1 grid showed differences of less than 0.2~meV per C atom in the energy differences between different structures. 

\section{\label{results}Results and discussions}

\subsection{Optimisation of graphene growth}

\begin{figure}[!hbt]
	\centering
	\includegraphics[width=7.8cm]{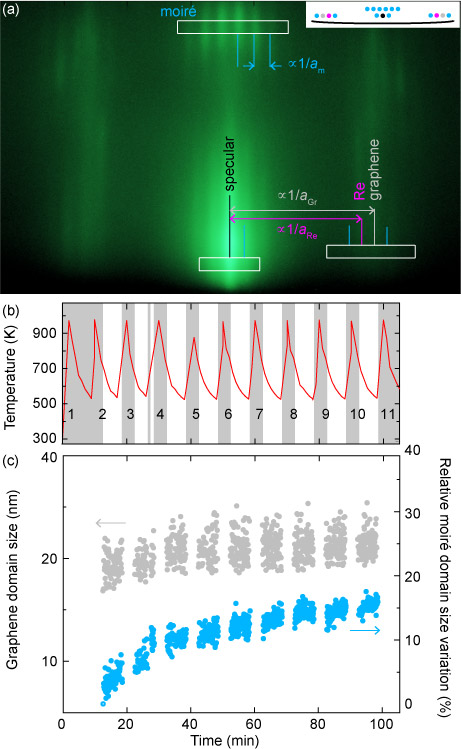}
	\caption{(a) RHEED pattern of graphene-covered Re(0001) along the [01$\bar{1}$0] incident azimuth. Coloured lines indicate the specular (black), Re (pink), graphene (grey) and moir\'e (blue) streaks (lines are superimposed only on the right side of the pattern). The distance between neighbour streaks is inversely proportional to the lattice parameter. Regions of interest are indicated with white rectangles. Inset shows a schematic top view of the reciprocal space with all observed rods and a cut of the Ewald's sphere (black line). (b) Sample temperature as a function of time, during the growth process consisting of 11 temperature cycles. The first and eleventh cycles were not included in our structural analysis, and the RHEED patterns falling in grey rectangles were excluded from the analysis \cite{SM}. (c) Sizes of the structurally coherent domains in graphene and the moir\'e, as deduced by RHEED analysis as a function of time.}
	\label{fig1}
\end{figure}

To selectively prepare graphene we used $P_\mathrm{C_2H_4}$ = 5$\cdot$10$^{-7}$~mbar, 10 temperature cycles up to 970~K [Fig.~\ref{fig1}(b)]. Such a temperature was found optimal for reaching the highest graphene structural quality, and is a limit above which graphene transforms into the surface carbide. Figure~\ref{fig1}(a) shows a typical RHEED pattern of graphene-covered Re(0001). We observe streaks related to the average lattice structure of Re, graphene, and the moir\'e superlattice, the latter being a (quasi)coincidence lattice arising from the lattice mismatch between graphene and the Re(0001) substrate \cite{miniussi2011,tonnoir2013}.

To characterise the structure, a sequence of RHEED patterns was acquired as a function of growth time. Distinct regions of interest were selected in the RHEED patterns [Fig.~\ref{fig1}(a)] to specifically address the graphene lattice and the moir\'e contributions.

In our analysis, the Re lattice parameter is assumed constant, unaltered by the presence of graphene (local strain possibly induced by graphene should not alter the average Re lattice parameter) or few 100~K changes of temperature during the temperature cycles (which should induce 0.3\% lattice thermal contraction/expansion \cite{lide2008}, \textit{i.e.} smaller than the $\simeq$~1\% uncertainty on our extracted strain values).

No change in the average graphene lattice parameter (graphene streak position) is observed (see Fig.~S3 in the SM). Using the Re lattice parameter (streak position) as a reference, we find a 2.46~\AA\, lattice parameter, close to the literature values. Consistent with this analysis, the moir\'e superlattice is essentially constant, of 2.14~nm (see Fig.~S3 and related discussion in the SM).

Structural changes are more obvious in the size of the structurally coherent (\textit{i.e.}, scattering coherently in the RHEED experiments) graphene and moir\'e domains. Within a simple kinematic theory of electron diffraction the size of structurally-coherent domains (see discussion below) is inversely proportional to the width of the scattering rods perpendicular to the surface \cite{ichimiya2004}, \textit{i.e.} to the FWHM of the RHEED streaks.  

The FWHM of the Re streak (not shown) is unchanged from cycle 2 to cycle 10 and amounts to $\simeq$ 0.40~\AA$^{-1}$. Assuming that Re is structurally coherent over a size of 100~nm, the average size of the Re(0001) terraces, we expect a FWHM $\simeq 2\pi/\mathrm{size}$ = 0.006~\AA$^{-1}$. The observed larger FWHM is dominated by instrumental effects, especially the coherence length of the electron beam (see discussion in the SM), which is about 20~nm in the first Laue zone, where the modulus of the out-of-plane scattering vector is the shortest [close to the bottom horizontal in Fig.~\ref{fig1}(a)]. By analysing the Re streak we extract the contribution of these instrumental effects, which also intervene in the FWHM of the graphene and moir\'e streaks.

Figure~\ref{fig1}(c) shows the size of structurally-coherent graphene domains, extracted from the FWHM of the corresponding streaks after the contribution of instrumental effects has been removed. This size first increases by about 20\% from cycle 2 (19~nm) to 4 (22~nm). From cycle 4, the coherent domain size saturates. As shown below, this saturation is inherent to the measurement itself, and due to the limited coherence of the electron beam in the first Laue zone, which is precisely of the order of 20~nm.

\begin{figure}[!hbt]
	\centering
	\includegraphics[width=6.49cm]{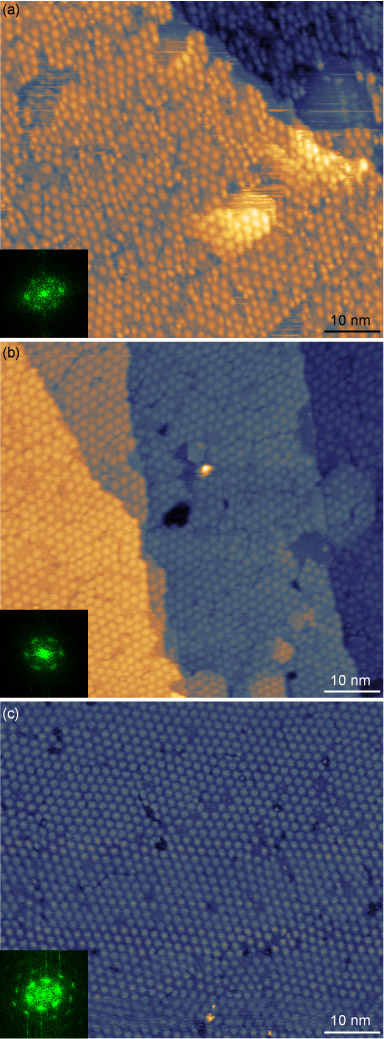}
	\caption{STM images (4~nA / 2~nA / 1~nA - 0.01~V / 1~V / 0.6~V) of graphene-covered Re(0001) after  one (a), two (b), and 10 (c) temperature cycles. The corresponding Fourier transforms are shown as insets.}
	\label{fig3}
\end{figure}

We turn to the analysis of the FWHM of the moir\'e streaks. Such streaks are found in the first Laue zone around the graphene and Re streaks, but there, their low intensity (hence signal-to-noise ratio) prevents from identifying any trend such as the one observed for the graphene streaks, because of the strong dispersion of the data. This is why we also analysed other moir\'e streaks, positioned above the specular streak in Fig.~\ref{fig1}(a), in the second Laue zone. For these streaks the FWHM is globally smaller due to an effectively larger coherence length of the electron beam. This means that we cannot extract absolute values of the domain sizes here (this would require to assess the contribution of instrumental effects using, \textit{e.g.} a Re streak in this particular region of reciprocal space, which we cannot do). Nevertheless, we can use relative units to accurately describe the variation of their FWHM. Figure~\ref{fig1}(c) reveals a clear increase of the moir\'e domain size from cycle 2 to 10, of about 15\%. While the increase is stronger for the first 3-4 cycles, no saturation such as the one observed for the graphene streaks is observed here.

Our RHEED observations are consistent with LEED measurements reported in Ref.~\onlinecite{miniussi2014}, where the moir\'e diffraction spots were found to increase in intensity at each temperature cycle.

Within the simple kinematic theory, the broadening of the graphene and moir\'e streaks/spots (and the reduction of their intensity) is due to the finite size of the structurally-coherent domains. The expression "structurally-coherent domains" refers to either actual graphene flakes, or regions within an extended layer where the atomic lattice is essentially undistorted -- in both cases, the domains that produce the constructive interferences at the origin of the diffraction streaks. To tell which case is relevant here, we use qualitative arguments provided by an STM analysis.

Figures~\ref{fig3}(a,b) show that the graphene coverage increases only marginally between the first and second cycle. To understand why this is so, we note that after the first temperature cycle, the sample has been exposed to a large dose of typically 100~L of ethylene at temperatures above 700~K. Since at these elevated temperature, ethylene molecules should be efficiently cracked by the Re(0001) surface, the observed close-to-100\% graphene coverage is not a surprise. This suggests that the effect limiting the structural coherency of graphene should be lattice distortions (strains).

Although the STM images shown in Fig.~\ref{fig3} are not atomically-resolved (hence they do not directly reveal strains), they reveal the moir\'e pattern. Strains in the carbon lattice reflect in the moir\'e lattice, where they are amplified \cite{coraux2008}, here by a factor of about 10. Strong distortions are indeed observed in the STM images for the first two cycles [Figs.~\ref{fig3}(a,b); even more so after the first cycle, see Fig.~\ref{fig3}(a)]: the moiré hills (regions of larger apparent height) are positioned onto a distorted lattice and they occasionally or often appear elongated or compressed. In addition to these deformations, defect occurring at local scale (and as such not liable to broaden the diffraction streaks) are also observed. In particular, so-called vacant moir\'e hills, corresponding to stacking faults in either the metal or graphene lattice \cite{artaud2020,pochet2017}, are found. As the number of temperature cycles increases, the amount of distortions and local defects is reduced [Figs.~\ref{fig3}(a-c)]. Consistent with these real-space observations and with the RHEED data, the Fourier transforms of the STM images (insets in Fig.~\ref{fig3}) exhibit better-defined and more numerous harmonics as the number of cycles increases. The progressive ordering of the graphene and moir\'e (super)lattices as a function of the number of cycles may occur \textit{via} the rotation of less-energetically-favourable domains, the incorporation of carbon nanoclusters \cite{artaud2018}, and the release of heterogeneous strain in graphene (for instance through local sliding of the graphene lattice with respect to that of the substrate or through the healing of point defects such as vacancies \cite{blanc2013}).
  
Obviously, increasing the number of temperature cycles has positive effects on the quality of graphene grown on Re(0001). Increasing the time of the final annealing, even at lower temperature, does not have the same effects, and instead transforms graphene into a surface rhenium carbide.

\subsection{STM and RHEED observations of the rhenium surface carbide}

\begin{figure}[!hbt]
	\centering
	\includegraphics[width=6.44cm]{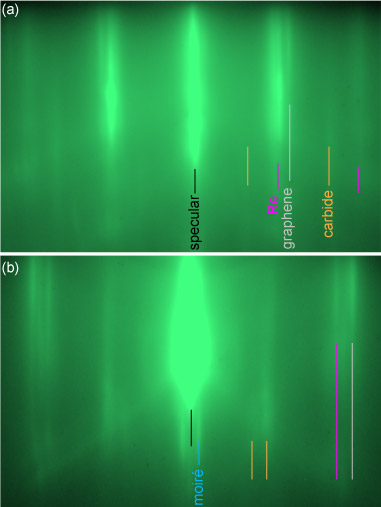}
	\caption{RHEED patterns of coexisting graphene and surface carbide phases along the (a) [11$\bar{2}$0] and (b) [01$\bar{1}$0] incident azimuths. Vertical coloured lines indicate the specular (black), Re (pink), graphene (grey), moir\'e (blue) and surface carbide (orange) streaks.}
	\label{figRHEEDcarbide}
\end{figure}

To promote the formation of the surface carbide, we found that decreasing $P_\mathrm{C_2H_4}$ to 1$\cdot$10$^{-7}$~mbar during the temperature cycles, and decreasing the temperature of the annealing steps to 870~K are well suited. After four temperature cycles and a fifth one in the absence of C$_2$H$_4$, the sample was finally annealed for 1~h at 870~K.

In the RHEED patterns, additional streaks are observed compared to the previous sample (where only graphene was formed). Along two different azimuths, two new characteristic streaks are observed (Fig.~\ref{figRHEEDcarbide}). These streaks correspond to the intersection of the Ewald's sphere with scattering rods extending perpendicular to the surface. The footprint of these scattering rods in the surface plane was observed in previous LEED data \cite{miniussi2014,zimmer1972} (see Fig.~S4 in the SM). These LEED patterns were attributed to a surface rhenium carbide with (7 $\times$ $\sqrt{\mathrm{19}}$) unit cell, \textit{i.e.} with lower symmetry than the Re(0001) surface, and appearing in the form of domains with different crystallographic orientations.

\begin{figure*}[!hbt]
	\centering
	\includegraphics[width=14.15cm]{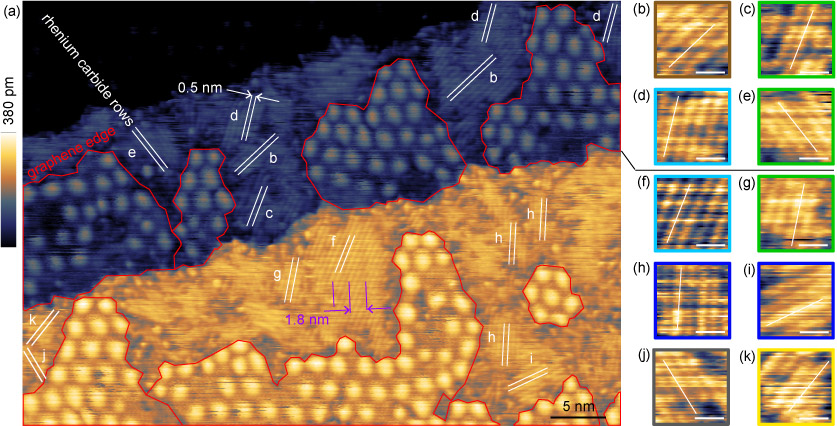}
	\caption{(a) STM images (2~nA, 1~V) of a Re(0001) surface with coexisting graphene and surface carbide. Carbide variants with different orientations are marked with letters (b-k). Purple lines highlight the modulation of apparent height observed on certain variants of the carbide phase. (b-k) Close-up views (scale bars, 1~nm) of the ten (out of 12 possible; \textit{b}, \textit{d}, \textit{h} are observed several times) carbide phase variants, each with a distinct crystallographic orientation, identified in (a), where (b-e) are the variants identified in the lower terrace, and (f-k) are the variants identified in the upper terrace. Equivalent variants (due to the surface symmetry or on distinct terraces) are signaled with a single colour of the frame around the subfigures.}
	\label{fig4}
\end{figure*}

Figure~\ref{fig4}(a) shows a representative STM image of the surface. Besides the graphene patches (easily recognized from their moir\'e patterns), isolated clusters assigned to carbon clusters of well-defined size \cite{gall1987,artaud2020}, and bunches of parallel lines separated by 0.5~nm are observed. The latter structure covers about 30\% of the surface in the preparation conditions we have chosen, and accounts for our RHEED observations (see below). We hence ascribe it to the surface rhenium carbide.

\subsection{Orientation variants of the surface carbide and symmetry considerations}

The STM images make it obvious that the surface carbide has several well-defined crystallographic orientations [Fig.~\ref{fig4}(a)]. This is consistent with previous LEED observations \cite{zimmer1972,miniussi2014} (also see Figs.~S4,S5 in the SM), which were interpreted by invoking the coexistence of six kinds of domains \cite{miniussi2014}. 

\begin{figure*}[!hbt]
	\centering
	\includegraphics[width=12.647cm]{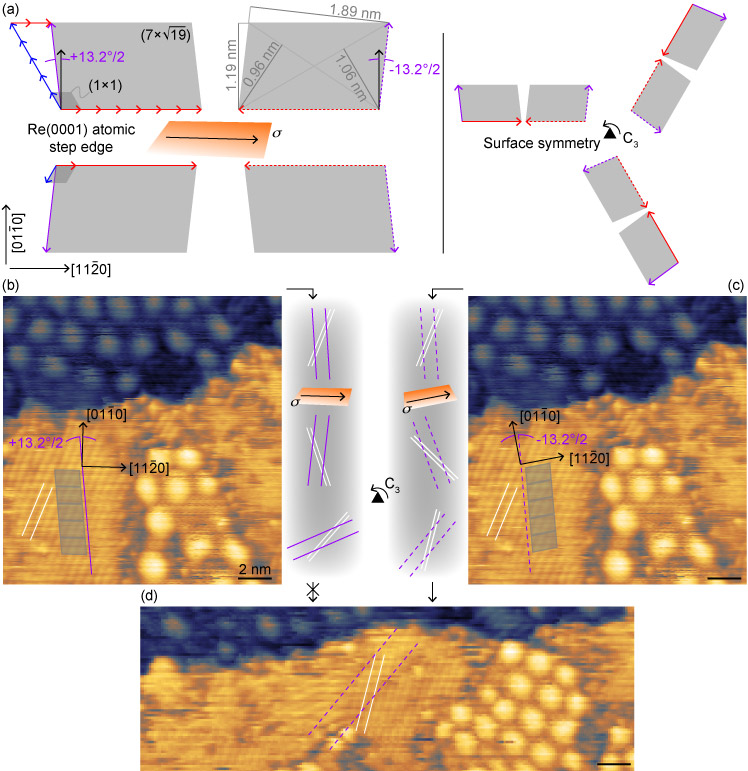}
	\caption{Symmetry of the carbide's lattice in its different variants. (a) On given Re(0001) terrace, there are two kinds (solid and dotted arrows) of (7 $\times$ $\sqrt{\mathrm{19}}$) unit cells [large grey parallelograms constructed from the Re(0001) unit cell shown with a small grey (1$\times$1) rhombus], each reflected with respect to the corresponding unit cell in the neighbour Re(0001) terrace (left). Each of the two kinds of unit cells on each terrace has two other equivalent unit cells, according to the C$_3$ symmetry of the surface (right). (b,c) Close-up views of Fig.~\ref{fig4}(a). The (7 $\times$ $\sqrt{\mathrm{19}}$) unit cells (grey parallelograms) fit in between two purple lines that mark the apparent modulation along the line pattern of the carbide. Two possible variants are considered in (b) and (c), each corresponding to a different orientation of the [11$\bar{2}$0] Re direction. (d) STM image (2~nA, 1~V) of a carbide  variant on a terrace adjacent to the one of the carbide in (b,c). At the center, the carbide pattern is mirrored with respect to a $\sigma$ plane (step edge) containing the [11$\bar{2}$0] direction and then rotated about a C$_3$ symmetry axis (surface symmetry). The resulting patterns match for (c,d).} 
	\label{fig5}
\end{figure*}

The possible (7 $\times$ $\sqrt{\mathrm{19}}$) unit cells of each of these six domains are sketched in Fig.~\ref{fig5}(a) (right pannel) for one specific terrace. They consist of three pairs, related to one another by 120$^\circ$-rotations about the C$_3$ symmetry axis standing perpendicular to the surface. Since there are two possible kinds of terraces on Re(0001) according to its hexagonal compact stacking, a total of 12 different kinds of carbide domains is expected in STM. On the left pannel of Fig.~\ref{fig5}(a), one kind of pair is sketched on either sides of a Re(0001) atomic step edge. The symmetry operation relating the unit cells on the left and right terraces is a mirror ($\sigma$) plane containing a $\left\langle 11\bar{2}0\right\rangle$ direction of Re(0001). To complete the description, we turn to the two unit cells of each pair (on each terrace): they are related to one another by a mirror operation, about a plane containing a $\left\langle 01\bar{1}0\right\rangle$ direction of Re(0001). While one of them has unit cell vectors of $7\vec{a}_{Re,1}$ and $2\vec{a}_{Re,1}+5\vec{a}_{Re,2}$ [solid arrows in Fig.~\ref{fig5}(a), left], for the other the unit cell vectors are $-7\vec{a}_{Re,1}$ and $3\vec{a}_{Re,1}+5\vec{a}_{Re,2}$ [dotted arrows in Fig.~\ref{fig5}(a), left]. The vectors $2\vec{a}_{Re,1}+5\vec{a}_{Re,2}$ and $3\vec{a}_{Re,1}+5\vec{a}_{Re,2}$ form a $\pm 13.2^\circ/2$ angle with the $[01\bar{1}0]$ direction \cite{noteonangle}.

We wish to determine the unit cells [among those presented in Fig.~\ref{fig5}(a)] to the variants identified in Fig.~\ref{fig4}. To do so, we must first unambiguously identify the above-mentioned mirror planes, \textit{i.e.} the substrate crystallographic directions ([11$\bar{2}$0] and [01$\bar{1}$0]) along which a mirror operation transforms the unit cell of a domain into a the unit cell of another kind domain on the same Re(0001) terrace or on an adjacent terrace. The bare Re(0001) surface is however not directly accessible here, hence it is not possible to directly identify these directions with atomic resolution imaging. The substrate's crystallographic directions cannot be identified indirectly either from the moir\'e pattern crystallographic directions in the neighbour graphene patches, since their orientation is known to be significantly scattered (it actually amplifies the rotations of the graphene lattice with respect to the metal lattice by a factor 10 \cite{coraux2008}).

More useful is the orientation of a feature that we have overlooked so far in the STM images. A modulation is observed (highlighted with purple lines) in the apparent height of the linear pattern of the surface carbide [Fig.~\ref{fig5}(b), corresponding to a zoom on the variant labeled ``f'' in Fig.~\ref{fig4}(a)]. The periodicity of this modulation is 1.8~nm, which is very close to the 1.89~nm value for the long period of a tiling with the (7 $\times$ $\sqrt{\mathrm{19}}$) unit cell [Fig.~\ref{fig5}(a)]. As shown in Figs.~\ref{fig5}(a,b,c), this match suggests that the $[01\bar{1}0]$ direction of Re(0001) is either $+13.2^\circ/2$ or $-13.2^\circ/2$ off the direction of the purple line, \textit{i.e.} the purple line aligns the unit cell vector $2\vec{a}_{Re,1}+5\vec{a}_{Re,2}$ [(7 $\times$ $\sqrt{\mathrm{19}}$) unit cell with solid vectors in Fig.~\ref{fig5}(a)] or the unit cell vector $3\vec{a}_{Re,1}+5\vec{a}_{Re,2}$ [(7 $\times$ $\sqrt{\mathrm{19}}$) unit cell with doted vectors in Fig.~\ref{fig5}(a)].

To determine which of the two Re(0001) crystallographic orientations ($\pm$13.2$^\circ$/2) is the actual one, we compare the two rhenium carbide domains of kind ``f'' located on the two adjacent terraces (separated by an atomic substrate step edge) imaged in Figs.~\ref{fig5}(b,c). For each of the two Re(0001) crystallographic configurations ($\pm$ 13.2$^\circ$/2), we applied a mirror operation to the carbide's pattern, with a $\sigma$ mirror plane containing the [11$\bar{2}$0] Re(0001) direction. This operation ``transfers'' the orientation of the carbide's pattern to the other terrace. After this operation, only one configuration [Fig.~\ref{fig5}(c)] yields the same orientation (modulo 120$^\circ$) for the two domains, which hence identifies the actual [11$\bar{2}$0] Re(0001) direction. Note that this orientation matches the average orientation of the substrate step edge \cite{noteonazimuth}. Given the high temperature used to prepare the Re(0001) surface, this might not be accidental: such a step edge orientation is energetically more favourable.

Now, the carbide domains can be categorised in families, each with a specific kind of unit cell shown in Fig.~\ref{fig4}(a). Variants are grouped \textit{via} 120$^\circ$-rotations within a terrace, and \textit{via} mirror symmetries along the [11$\bar{2}$0] directions of the Re(0001) between terraces. Overall, we find six families of equivalent atomic arrangements. They are highlighted with different colors in Figs.~\ref{fig4}(b-k). 

\subsection{Possible atomic structures of the motif of the rhenium carbide}

\begin{figure}[!hbt]
	\centering
	\includegraphics[width=6.08cm]{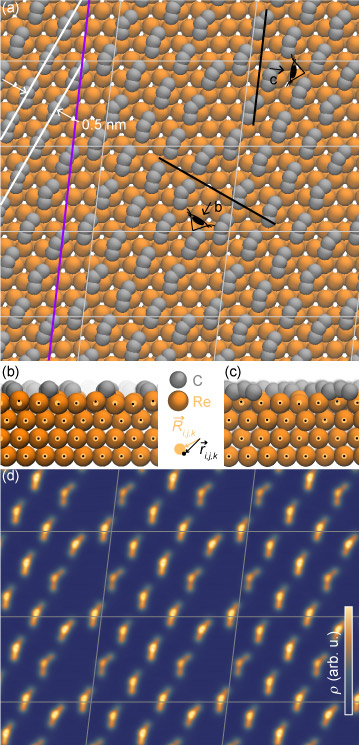}
	\caption{(a) Atomic details of the lattice motif of the second lowest-energy structure of a surface rhenium carbide variant labeled ``f'' in Fig.~\ref{fig4}(a), optimised by DFT calculations. (b,c) Cross-sections along the two directions marked in (a). The orange and black dots mark the positions $\vec{R}_{i,j,k}$ and $\vec{r}_{i,j,k}$ of Re atoms (labelled by indexes $i,j,k$) in an undistorted lattice and in the actual lattice, respectively. (d) Cut (0.6~\AA\; higher than the average height of topmost atoms) in the corresponding calculated electronic density ($\rho$) map.}
	\label{fig7}
\end{figure}


Beyond these symmetry considerations, the atomic details of the surface carbide remain essentially unknown. Although the STM images presented in Figs.~\ref{fig4},\ref{fig5} reveal an atomic-scale contrast, they do not directly unveil the atomic arrangement of rhenium carbide, but essentially reveal lines (0.5~nm separation) with a specific crystallographic orientation and a height modulation pattern (1.8~nm period) \cite{noteonresexp}. Former rhenium core level photoemission data together with DFT calculations \cite{miniussi2014} indicate that the lines actually consist of rows of specific kinds of trimers of C atoms. These trimers consist of C atoms bonded to so-called $fcc$ and $hcp$ hollow sites of Re(0001), sitting on top of a Re atom and a hollow site in the second Re layer respectively. The core level data also revealed a close-to-50\% weight of the C atoms on these two binding sites \cite{miniussi2014}.

We were able to produce four tentative structures compatible with all these experimental facts -- three have a unique trimer crystallographic orientation, one features trimers with two kinds of crystallographic orientations. All these structures have one half of the C atoms occupying $hcp$ sites (and the other half the $fcc$ sites). Their composition can be described as Re$_{11}$C$_{8}$, when considering the number of hollow sites of the Re(0001) surface that are occupied by C atoms. We optimised the atomic positions in these structures with the help of DFT calculations [Fig.~\ref{fig7}(a) and Fig.~S7 in the SM~\cite{SM}]. Obvious distortions are seen within each unit cell: on top and side views [Figs.~\ref{fig7}(a-c) and Fig.~S7 in the SM~\cite{SM}], the C trimers do not all look the same, and the Re atoms are displaced (by several 1\% both within the surface plane and perpendicular to it), most prominently in the topmost Re layer.

The DFT calculations show that all the structures are stable. We first focus on the optimised structure shown in Fig.~\ref{fig7}(a), where the $hcp$-$fcc$-$hcp$ trimers align along segments that span across three (7 $\times$ $\sqrt{\mathrm{19}}$) unit cells. These segments have exactly the same crystallographic orientation as those highlighted with white lines for variant labeled ``f'' in Fig.~\ref{fig4}(a). At the end of these segments, the atomic C density is locally lower, providing a possible interpretation for the observed 1.8~nm-periodic modulation of the apparent height highlighted with purple lines for this variant in Figs.~\ref{fig4}(a),\ref{fig5}(b-d). Why such (rather complex) atomic configurations are selected during growth is an open question, whose answer is beyond the scope of the present work: they should correspond to an energy minimum of the system, either a local or a global one (depending on kinetic hinderances), which to some extent mitigates elastic deformations in, \textit{e.g.} the substrate and the C trimers, repulsive C-C interactions mediated by the substrate \cite{artaud2018}, etc. Overall, the constant-height cut in the DFT-simulated electronic density map corresponding to this structure [Fig.~\ref{fig7}(d)] appears compatible~\cite{noteonres} with our STM observations of this variant.

The energy of the structure we just discussed is higher by 0.9~eV per unit cell than the one of the lowest-energy structures (Fig.~S6 in the SM~\cite{SM}). Altogether, the four structures that were optimised with DFT differ by 0.4 to 1.1~eV per unit cell. Expressed relative to the number of C atom in each unit cell, these energy differences only amount to few millielectronvolts, which are small values compared to thermal energy (few to several 10~meV) in the growth and STM imaging conditions. It is hence reasonable to expect that the different models we propose in Fig.~S6 in the SM~\cite{SM} actually each correspond to different kinds of variants observed for instance in Fig.~\ref{fig4}. Interestingly, the binding energy per C atom in the carbide structures we have considered are about 500~meV lower than that of C atoms in graphene/Re(0001), which hence is predicted to be the thermodynamically stable form of C. Beyond the scope of the present work and as already mentioned in the introduction, it would certainly be interesting to understand the influence of the C chemical potential at the surface of Re(0001), and in particular of the excess chemical potential with respect to the critical value corresponding to graphene and carbide nucleation. The relative stability of the different C phases may depend, at least quantitatively, on this excess value. Please also note that while our calculations are performed for a surface carbide structure extending to infinity (fully-periodic system), our experiments reveal variants in the form of domains with finite extension in the range of 5 to 10~nm [Fig.~\ref{fig4}(a)]. There, finite-size and boundary effects may alter the energy hierarchy between the different structures and phases we considered.

\section{Conclusion}

We have investigated the structure of graphene and a surface carbide grown on the (0001) surface of Re, using RHEED, STM, and DFT calculations. We find that the number of cycles used to grow graphene is a key parameter to improve its structural quality. On the contrary, long annealing times, that could give time for the system to heal its defects, do not have the desired effect. They promote the transformation of graphene into a surface carbide with (7$\times$$\sqrt{19}$) unit cell. We were able to identify the possible crystallographic orientations (variants) of two six-fold families of equivalent domains of the rhenium surface carbide. The carbide is highly anisotropic, consisting of parallel lines separated by 0.5~nm, and periodically modulated at the nanometer scale. We have proposed a realistic atomistic model of one of the carbide variants, whereby the observed lines consist of eight adjacent carbon trimers, in agreement with previous photoemission data. In a C-Re phase diagram, the rhenium carbide would represent a rather low-density C phase, in this sense intermediate between graphene and a dilute phase \cite{gall1987} in the Re bulk. We suggest that the surface rhenium carbide could serve as a template for the self-organisation of functional nanoclusters or molecules.
\vspace{5mm}

\section*{Acknowledgments}
This work was supported by the R\'egion Rhône Alpes (ARC6 program) and the Labex LANEF. Funding from the French National Research Agency under the 2DTransformers (OH-RISQUE program ANR-14-OHRI-0004) and ORGANI'SO (ANR-15-CE09-0017) projects is gratefully acknowledged. MP and DA are supported by the Natural Environment Research Council Grant No. NE/R000425/1. DFT calculations were performed on the Monsoon2 system, a collaborative facility supplied under the Joint Weather and Climate Research Programme, a strategic partnership between the UK Met Office and the Natural Environment Research Council.


\begin{thebibliography}{55}
\bibitem [{not({\natexlab{b}})}]{noteonazimuth}%
  \BibitemOpen
  \href@noop {} {} \bibinfo {note} {Also
  note that this assignment is consistent with the crystallographic orientation
  that can be deduced from a RHEED search for the main crystallographic
  directions as function of the azimuthal angle. In our experimental setup
  unfortunately, absolute measurements of this angle are not precise, due to a
  significant non-reproducible backlash in the azimuthal rotation axis. We can
  only assess the absolute azimuthal angle \textit{via} a visual inspection of
  the orientation of the sample with respect to the electron beam axis, which
  is no better than $\pm$10$^\circ$.}\BibitemShut {Stop}%
\bibitem [{not({\natexlab{c}})}]{noteonresexp}%
  \BibitemOpen
  \href@noop {} {} \bibinfo {note} {Due to the very
  demanding STM imaging conditions (the uneven surface, with a coexistence of
  graphene, carbide, and other adsorbed species, leads to frequent
  modifications of the STM probe apex), we are unable to collect higher
  resolution data than in Figs.~\ref{fig4},\ref{fig5}, or a broader set of data
  allowing to perform, for all domains observed in STM, the analysis we
  performed form domains of type ``f'' (Fig.~\ref{fig5}).}\BibitemShut {Stop}%
\bibitem [{not({\natexlab{d}})}]{noteonres}%
  \BibitemOpen
  \href@noop {} {} \bibinfo {note} {Please keep in mind that
  our calculations, which do not include the effect of the STM tip, are better
  resolved than the experimental STM images.}\BibitemShut {Stop}%
\end{thebibliography}

%

\end{document}